\documentclass[conference]{IEEEtran}
\IEEEoverridecommandlockouts
\usepackage{multirow}
\usepackage{cite}
\usepackage{amsmath,amssymb,amsfonts}

\usepackage{graphicx}
\usepackage{textcomp}
\usepackage{pifont}
\usepackage{xcolor}
\usepackage{pgfplots}
\pgfplotsset{compat=1.18} 
\usepackage[most]{tcolorbox}
\usepackage{listings}
\usepackage{tikz} 
\usepackage[table,xcdraw]{xcolor}
\newcommand*\circled[1]{\tikz[baseline=(char.base)]{
            \node[shape=circle,draw,inner sep=0.5pt] (char) {#1};}}

\usepackage{algorithm}
\usepackage{algpseudocode}
\usepackage{amsmath}
\usepackage{soul}
\usepackage{xcolor}
\usepackage{array}
\usepackage{booktabs}
\usepackage{tabularx}
\usepackage{tcolorbox}
\usepackage{xcolor}
\tcbuselibrary{breakable, skins, listings}

\usepackage{listings}
\usepackage{xcolor}
\definecolor{codegray}{gray}{0.95}

\lstdefinelanguage{PromptPython}{
    language=Python,
    keywords={import, from, class, def, return, True, False, None},
    keywordstyle=\color{blue}\bfseries,
    commentstyle=\color{gray}\ttfamily,
    stringstyle=\color{black},
    basicstyle=\ttfamily\footnotesize,
    identifierstyle=\color{black},
    morestring=[b]',
    morestring=[b]",
    showstringspaces=false,
    breaklines=true,
    breakatwhitespace=true,
    numberstyle=\tiny\color{gray},
    numbersep=6pt,
    rulecolor=\color{black},
    backgroundcolor=\color{codegray},
    captionpos=b
}

\usepackage{listings}
\lstdefinelanguage{Verilog}{
  keywords={module, endmodule, input, output, wire, reg, always, if, else, assign, begin, end, property, assert, disable, iff, endproperty, posedge, process, end process, or, and, then, elsif, case, always_ff, logic, typedef, enum, endcase, covergroup, coverpoint, endgroup},
  keywordstyle=\color{blue}\bfseries,
  morecomment=[l]{//},
  morecomment=[s]{/*}{*/},
  commentstyle=\color{gray}\ttfamily,
  morestring=[b]",
  stringstyle=\color{red},
  identifierstyle=\color{black},
 basicstyle=\ttfamily\footnotesize,
  sensitive=true,
  showstringspaces=false,
  numbers=left,
  numberstyle=\tiny\color{gray},
  stepnumber=1,
  numbersep=5pt,
  breaklines=true,
  captionpos=b
}

\def\BibTeX{{\rm B\kern-.05em{\sc i\kern-.025em b}\kern-.08em
    T\kern-.1667em\lower.7ex\hbox{E}\kern-.125emX}}
    
\begin{document}

\bstctlcite{IEEEexample:BSTcontrol}

\title{CITADEL: \underline{C}WE-Guided \underline{I}nsertion of Hardware \underline{T}rojans via \underline{A}nalysis of \underline{D}FG-\underline{E}nabled \underline{L}LMs \vspace{-0.5 em}} 

\author{
\IEEEauthorblockN{
Jayanth Thangellamudi\IEEEauthorrefmark{1},
Raghul Saravanan\IEEEauthorrefmark{1},
Sudipta Paria\IEEEauthorrefmark{2},
Swarup Bhunia\IEEEauthorrefmark{2}, and
Sai Manoj P D\IEEEauthorrefmark{1}
}
\IEEEauthorblockA{
\IEEEauthorrefmark{1}Department of Electrical and Computer Engineering, George Mason University, Fairfax, VA
}
\IEEEauthorblockA{
\IEEEauthorrefmark{2}Department of Electrical and Computer Engineering, University of Florida, Gainesville, FL\\
jthangel@gmu.edu, rsaravan@gmu.edu, sudiptaparia@ufl.edu, swarup@ece.ufl.edu, spudukot@gmu.edu
}
}

\maketitle

\begin{abstract}
The increasing sophistication of Hardware Trojans (HTs) and system-level vulnerabilities poses significant risks to modern integrated circuits. However, constructing realistic HT scenarios, remains a substantial burden: researchers must manually analyze complex RTL structures, identify plausible weaknesses, and craft stealthy, synthesizable insertions that preserve functional correctness. This paper introduces \textbf{CITADEL} (CWE-Guided Insertion of Trojans via Analysis of DFG-Enabled LLMs), a framework that leverages Large Language Models (LLMs) and Data Flow Graphs (DFGs) to automate CWE-grounded HT synthesis. CITADEL uses structured CWE semantics together with DFG-derived structural context to assist the user in identifying relevant vulnerabilities, localize the module surrounding the chosen insertion point, and perform intent-conditioned RTL modification. The framework produces minimal, synthesizable, and interface-preserving HTs with ultra-rare triggers. Experimental evaluation across diverse RTL designs demonstrates that all generated HTs are 100\% syntactically correct, remain undetectable under large-scale random simulation, and are functionally triggerable under their intended activation conditions. These results highlight CITADEL as a scalable and principled method for generating realistic HT benchmarks.
\end{abstract}


\section{Introduction}
The globalization of the semiconductor supply chain has fundamentally reshaped integrated circuit (IC) design and manufacturing. To remain competitive, design houses increasingly depend on third-party intellectual property (3PIP) blocks and fabrication at external foundries \cite{feng22, Ghosh23}. Establishing and maintaining advanced fabrication facilities is prohibitively expensive, often requiring investments of several billions of dollars \cite{TAINT}. As a result, outsourcing has become the industry standard, lowering costs and accelerating time-to-market. However, this distributed model also introduces significant security risks, since untrusted entities at different stages of the pipeline may alter the design \cite{tehranipoor2010survey}. Among these risks, Hardware Trojans (HTs) remain one of the most insidious threats: malicious circuit modifications that can remain dormant during validation only activate under rare conditions to compromise confidentiality, integrity, or availability \cite{saravanan2024emergence}.


To study and evaluate such threats, numerous HT insertion frameworks have been proposed; however, as discussed in Section \ref{sec:compare}, they suffer from fundamental limitations. Existing approaches span manual netlist edits, semi-automated FPGA-level perturbations, and probabilistic or ML-based insertion strategies \cite{ATTRITION,MIMIC,TrojanForge}. Despite this apparent diversity, most frameworks still rely on handcrafted HT patterns or require substantial manual effort,  limiting their scalability to the structural complexity of modern RTL designs. Furthermore, their lack of explicit awareness of module hierarchy and
data-flow relationships restricts the realism of the HTs they generate. Public benchmarks such as Trust-Hub \cite{trust} offer valuable case studies, but they remain narrow, manually curated examples that fail to capture the breadth of vulnerabilities present in modern systems. These limitations collectively motivate the need for a principled, scalable methodology capable of generating realistic HT scenarios directly grounded in standardized vulnerability taxonomies.

A natural foundation for enabling systematic HT construction is the Common Weakness Enumeration (CWE), a curated vulnerability catalog maintained by MITRE \cite{mitre_cwe}. CWE encompasses both software and hardware weaknesses, with hardware entries describing issues such as improper access control, unprotected communication paths, and insecure state machines. By providing a structured classification of vulnerabilities, CWE serves as an important foundation for hardware security research. Aligning HT scenarios with CWE entries ensures coverage across a broad threat space, comparability across research efforts, and reproducibility anchored in a widely recognized standard. Despite these advantages, applying CWE directly to hardware designs remains challenging, as mapping abstract vulnerability categories to concrete RTL constructs and signal dependencies is still largely manual and expertise-driven, which limits scalability.

Advances in Large Language Models (LLMs) open new opportunities to address this challenge \cite{xu2025largelanguagemodelsllms}. LLMs have demonstrated strong capabilities in semantic reasoning, program analysis, and code generation across structured and unstructured data \cite{blocklove2024evaluating,llm_survey}. In hardware design, they have already been applied to tasks such as RTL synthesis, verification, and bug detection \cite{ALSAQER20252392}, showing their ability to interpret domain-specific representations. Extending these capabilities to hardware security enables a new research direction: LLMs can parse Data Flow Graphs (DFGs) derived from hardware designs, reason about their structure, and align design-level features with CWE descriptions. This enables not only the identification of vulnerabilities but also the generation of CWE-guided HT scenarios that are realistic, stealthy, and reproducible. By reframing HT analysis as a reasoning and generation task, LLMs provide a pathway to automate what has historically been manual, error-prone, and limited in scope.

The cardinal contributions of this paper are summarized as follows:
\begin{itemize}
    \item We present the first HT-insertion framework that integrates DFGs with the CWE hardware taxonomy to generate vulnerability-grounded HT scenarios.
    
    \item We develop an LLM-assisted CWE applicability analysis that interprets DFG structural cues to provide rationale for vulnerability relevance, supporting the user in selecting suitable CWE mechanisms for HT insertion.
    
    \item We propose an intent-conditioned RTL modification pipeline in which the LLM, guided by DFG context and strict prompt constraints, synthesizes minimal, synthesizable, and module-localized HTs aligned with CWE.
    
    \item We validate the framework on diverse RTL designs through syntax checking, formal trigger reachability, and large-scale random simulation, demonstrating correctness, reachability, and strong stealthiness.
\end{itemize}

\section{Background and related work}
\subsection{Hardware Trojans (HTs)}
\begin{figure}[h]
    \centering
    \includegraphics[width=0.8\linewidth]{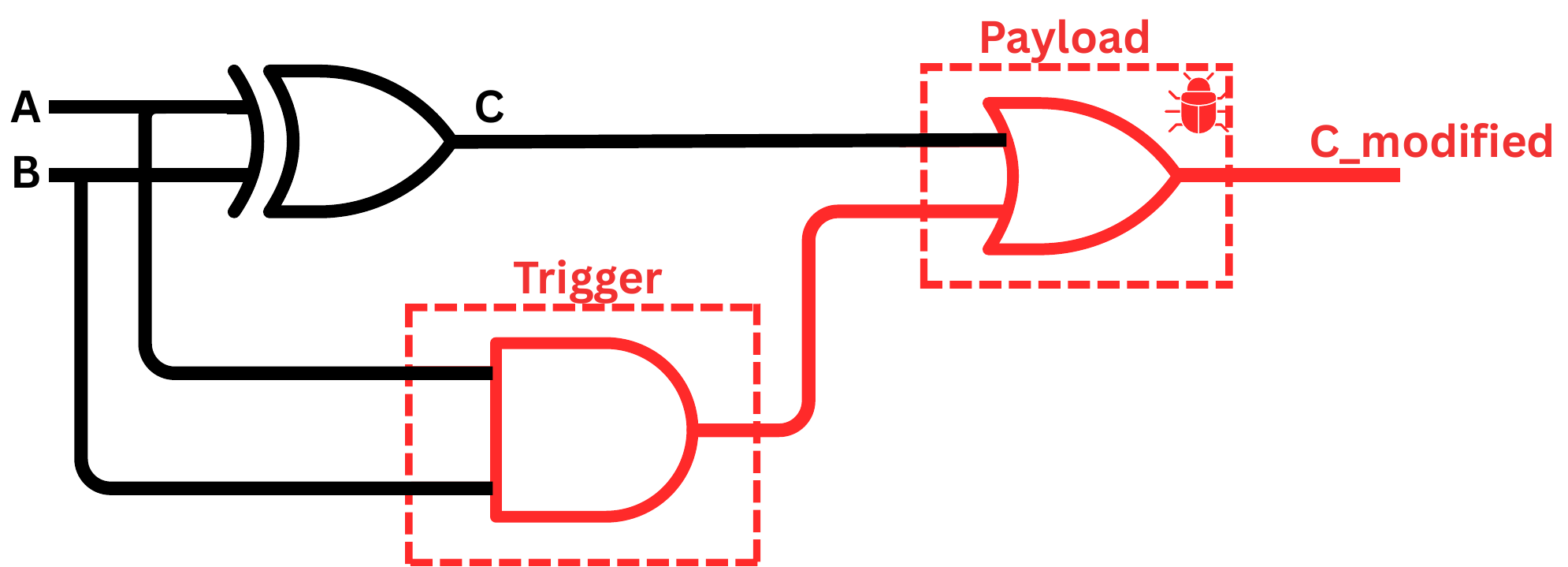}
    \caption{Combinational HT: Rare-event trigger with output forcing payload}
    \label{fig:HT}
    \vspace{-1em}
\end{figure}

HTs are malicious modifications to integrated circuits that alter functionality, leak sensitive information, or degrade reliability \cite{TAINT, swarup, saravanan2026automatedframeworkgeneratingstealthy}. They are typically characterised by a trigger that activates the malicious behaviour under rare conditions and a payload that executes the intended compromise. Even a small perturbation, such as the additional trigger and payload logic shown in Fig.~\ref{fig:HT} can silently override the intended output without affecting normal behaviour. HTs may be inserted at various stages of the design and manufacturing pipeline \cite{swarup}, from RTL development to fabrication, and are often designed to evade conventional testing by minimizing their impact on power, area, and performance. Because such HTs are intentionally stealthy and structurally subtle, evaluating detection mechanisms requires realistic HT examples; but constructing them manually is labour-intensive, design-specific, and difficult to scale. Recent advances in LLMs offer a promising direction toward addressing this gap.

\subsection{LLMs in Hardware Security}

LLMs have recently gained traction in security applications, from software vulnerability discovery to adversarial testing. In the hardware domain, LLMs have been explored for RTL code generation\cite{llm_survey,thakur2023benchmarking,rtlcoder, verirag, dnt }, bug detection\cite{divas,ahmad,lambda}, assertion synthesis\cite{assertllm,lasso,ip_cots}, and functional verification~\cite{menon2025vert}. These advances underscore the potential of LLMs to generalize across diverse hardware workflows. However, most existing approaches operate at the surface level without explicit structural awareness, limiting their effectiveness for security-critical applications. DFGs provide a natural substrate for bridging this gap, as they capture signal dependencies and module interactions essential for identifying vulnerable points~\cite{gnn4tj}. Leveraging LLMs with DFG context enables HT scenarios that are both automated and structurally consistent.

\begin{table}[h]
\centering
\caption{Comparison of HT insertion frameworks.}
\label{tab:citadel-compare}
\renewcommand{\arraystretch}{1.1}
\setlength{\tabcolsep}{2pt}

\begin{tabular}{|p{1.8cm}|p{2.5cm}|p{1.4cm}|p{1.4cm}|p{1.2cm}|}
\hline
\textbf{Framework} & \textbf{Insertion Strategy} & \textbf{Abstraction Level} & \textbf{Automation} & \textbf{Structural Awareness} \\
\hline
HAL~\cite{HAL} & Reverse-engineered net edits & Gate-level & \ding{55} & Partial \\
\hline
TAINT~\cite{TAINT} & FPGA LUT/FF/BRAM injection & RTL / Gate & Partial & \ding{55} \\
\hline
TRIT~\cite{TRIT} & Probability-based trigger nets & RTL & \ding{55} & \ding{55} \\
\hline
MIMIC~\cite{MIMIC} & ML-based HT generation & Gate-level & \ding{51} (ML) & \ding{55} \\
\hline
ATTRITION~\cite{ATTRITION} & RL-guided rare-net insertion & RTL / Gate & \ding{51} (RL) & \ding{55} \\
\hline
TrojanForge~\cite{TrojanForge} & GAN-based adversarial edits & Gate-level & \ding{51} (GAN) & \ding{55} \\
\hline
GHOST~\cite{ghost} & Prompt-based LLM edits & RTL & \ding{51} (LLM) & \ding{55} \\
\hline
\textbf{CITADEL (This Work)} & DFG-guided reasoning with LLM & RTL (DFG) & \ding{51} (LLM) & \ding{51} \\
\hline
\end{tabular}
\vspace{-1 em}
\end{table}

\subsection{Comparison of HT Insertion Frameworks}
\label{sec:compare}



Prior HT insertion frameworks vary widely in methodology, but most share a common limitation: they do not incorporate an explicit structural understanding of the design when generating HTs. Early tools such as \cite{HAL} and \cite{TAINT} rely on manual or template-driven edits, offering limited insight into how signal relationships or module boundaries influence HT behaviour. Later approaches, including \cite{TRIT}\cite{MIMIC},\cite{ATTRITION} and \cite{TrojanForge} introduced probabilistic, ML, RL, or GAN-based strategies to automate parts of the insertion process; the resulting HTs are often generated without awareness of internal dataflow dependencies or hierarchical semantics. Recent LLM-based work like \cite{ghost} demonstrates the feasibility of LLM-generated HTs, its approach is purely text-based and omits structural cues from the design’s DFG or control/data dependencies, leading to HT placements that may be syntactically valid but architecturally inconsistent. As summarised in Table~\ref{tab:citadel-compare}, existing methods typically lack \textit{(i) fine-grained structural context, (ii) vulnerability-grounded insertion logic, or (iii) mechanisms to ensure that the modification remains consistent with the RTL’s operational semantics.}

\textbf{CITADEL} overcomes these limitations by coupling LLM reasoning with explicit DFG-derived structure and CWE semantics. By exposing the LLM to instance hierarchy, term-node connectivity, and source–destination dataflow relations, CITADEL performs HT insertion that is structurally coherent, vulnerability-aligned, and minimally invasive, representing the first approach that systematically grounds HT synthesis in both design semantics and standardised vulnerability taxonomies.

\begin{figure}[!b]
    \centering
    \vspace{-1em}
    \includegraphics[width=0.95\columnwidth]{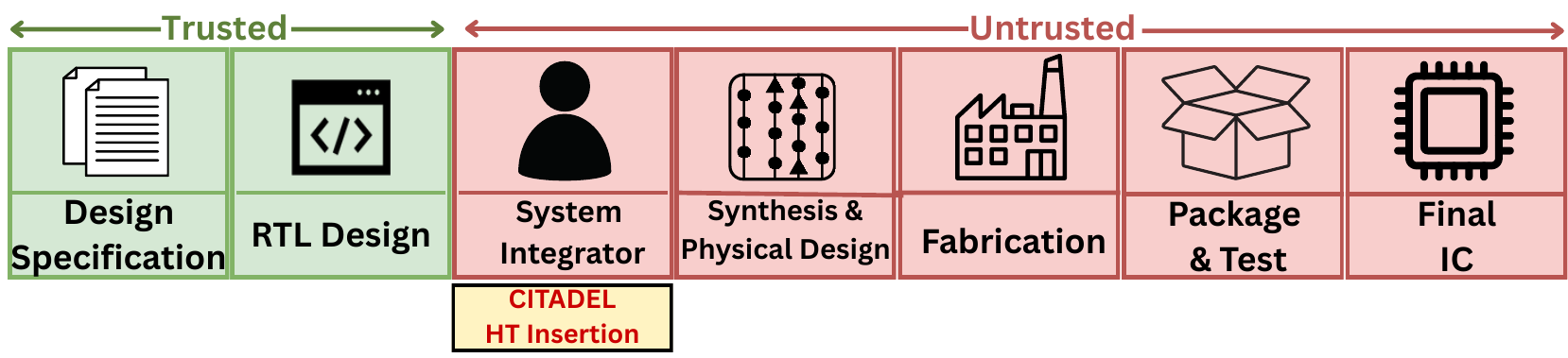}
    \vspace{-1em}
    \caption{Overview of the IC design flow and CITADEL-based HT insertion.}
    \label{fig:threat}
\end{figure}

\begin{figure*}[h]
    \centering
    \includegraphics[width=0.9\linewidth]{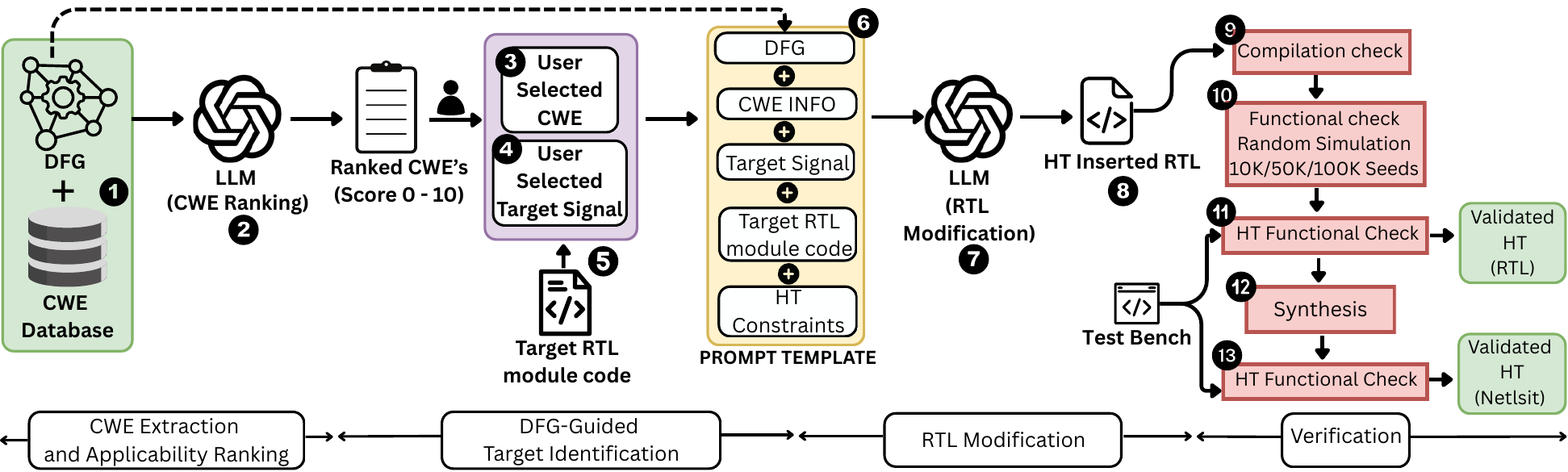}
    \caption{CITADEL framework overview. The DFG and CWE database are jointly processed by an LLM to rank CWE applicability. Based on user-selected CWE and target signal, a prompt template is constructed incorporating DFG context, CWE information, and HT constraints. The LLM then generates modified RTL with inserted HTs using the target module as reference. The output undergoes three-stage verification: syntax checking, stealthiness evaluation through large-scale random simulation, and Functional checking, producing validated HT RTL.}
    \label{fig:CITADEL}
    \vspace{-1em}
\end{figure*}

\subsection{Threat Model}


To situate CITADEL, we adopt a supply-chain threat model as illustrated in Figure \ref{fig:threat}  in which the RTL designer and fabrication foundry are trusted, while the system integration vendor is potentially adversarial as similar to threat model of \cite{threatmodel}. This adversary can modify RTL prior to synthesis but has no access to physical design stages, restricting attacks to synthesizable, functionally correct RTL-level insertions. HT logic must therefore be constructed solely from existing internal signals and remain consistent with the module’s structural and dataflow dependencies as exposed by the DFG. This setting reflects realistic risks in outsourced integration workflows and motivates CITADEL’s focus on generating structurally coherent, CWE-aligned RTL HTs for evaluating detection and defense mechanisms.

\section{Proposed Framework}

This section presents CITADEL, a CWE-guided HT insertion framework powered by LLMs which integrates three core capabilities: (1) structural analysis using DFGs, (2) vulnerability grounding via CWE semantics, and (3) intent-conditioned RTL modification through an LLM. These components collectively enable controlled HT synthesis that conforms to design semantics while maintaining dormant, ultra-rare activation behavior. Figure~\ref{fig:CITADEL} outlines the end-to-end pipeline, and the following subsections describe each stage in detail.

\subsection{Overview of CITADEL}
As shown in Fig.~\ref{fig:CITADEL}, CITADEL operates as a sequential pipeline that transforms CWE knowledge and DFG structure into validated HT-inserted RTL. The process begins by parsing the CWE database and the textual DFG \circled{1}, extracting vulnerability semantics and hierarchical dataflow context. These are supplied to the LLM \circled{2}, which generates a ranked list of CWE candidates applicable to the design. After the user selects a CWE and a target signal \circled{3},\circled{4}, CITADEL consolidates all conditioning information, CWE semantics, DFG context, target module code, and HT-insertion constraints, into a specialized prompt template \circled{6}. The LLM then produces a modified version of the target module containing a CWE-aligned HT \circled{7}, yielding HT-inserted RTL \circled{8}. This RTL undergoes a multi-stage verification flow \circled{9}–\circled{13},  after which CITADEL outputs fully validated HT-inserted code.

\subsection{CWE Extraction and Applicability Ranking}
CITADEL begins by ingesting two primary inputs: the MITRE CWE dataset and the DFG generated by dataflow analysis tool, as indicated by \circled{1} in Figure \ref{fig:CITADEL}.

\subsubsection{CWE Parsing and Structuring}
CITADEL begins by ingesting the CWE CSV repository \cite{mitre_cwe} as shown in Figure \ref{fig:CITADEL} \circled{1}, which includes a standardized subset of hardware-relevant weakness classes within the broader taxonomy. From the full dataset, only four fields are retained, CWE-ID, Name, Description, and Extended Description, and are normalized into a structured list of dictionaries. This machine-readable representation grounds subsequent LLM reasoning in the official taxonomy. The Name serves as a fixed identifier, the Description captures core vulnerability semantics, and the Extended Description details operational preconditions and failure modes. Together, these fields collectively enable the LLM to (a) map weaknesses to structural elements in the DFG, (b) infer valid trigger–payload patterns, and (c) generate RTL modifications consistent with the CWE semantics.



\subsubsection{DFG Parsing and Hierarchical Mapping}
Next, CITADEL \circled{1} parses the textual DFG to obtain a machine-readable representation of the design’s hierarchy and dataflow. The parser extracts:
\begin{itemize}
    \item \textbf{Instance-to-module mappings}, derived from tuples of the form (full.instance.path, 'module'), identifying where each instantiated component originates.
    \item \textbf{Term nodes}, corresponding to leaf-level signal endpoints (e.g., \verb|aes_cipher_top.u0.wo_0|), each tagged with its owning module inferred by prefix-matching against the instance map.
    \item \textit{Bind edges} extracted from Bind statements, encoding explicit source$\to$destination dataflow relationships.
\end{itemize}

The resulting DFG structure (nodes, edges, instance map) is serialized as compact JSON and reused across the pipeline. It provides the LLM with design-specific constraints during CWE scoring and later enables precise localization of the module surrounding the user-selected signal. Without this structured view, the LLM would produce superficial, design-agnostic results, leading to unrealistic or easily detectable HT.

\subsubsection{LLM-Based CWE Applicability Ranking}
\label{sec:cwe-ranking}

With both CWE entries and the parsed DFG available \circled{2}, the LLM assigns each weakness an applicability score from 0 (negligible relevance) to 10 (high structural compatibility). Scores are inferred from DFG-derived features such as signal criticality, dataflow position, control dependence patterns, fan-in/fan-out characteristics, and proximity to sensitive computation. A concise rationale accompanies each score, explaining the structural cues influencing the evaluation.

Results are returned in a standardized Markdown table format:
\texttt{| CWE-ID | Name | Score | Rationale |}.

The ranked list, accompanied by rationales, is presented to the user for manual selection. This hybrid human-in-the-loop approach ensures that the chosen weakness aligns with both the automated structural analysis and the researcher’s intent, while preserving transparency into the LLM’s decision-making process.

\subsection{DFG-Guided Target Identification} 



Once a CWE is selected \circled{3}, the user specifies the exact signal to instrument \circled{4} (e.g.,\texttt{aes\_cipher\_top.u0.wo\_0}). CITADEL resolves the module containing this signal by performing a deterministic longest-prefix match over the hierarchical instance map extracted from the DFG. This allows the framework to localize the precise RTL module and recover the structural neighborhood of the signal via its associated term nodes and bind edges. A key advantage of this DFG-based lookup is scalability, instead of exposing the LLM to an entire multi-module RTL design, the DFG provides a compact structural index, enabling CITADEL to request only the single module relevant to the selected signal. This ensures that HT insertion remains targeted, efficient, and structurally coherent even for designs with thousands of lines of RTL.


\subsection{Intent-Conditioned RTL Modification Using LLMs}

\begin{figure}[!ht]
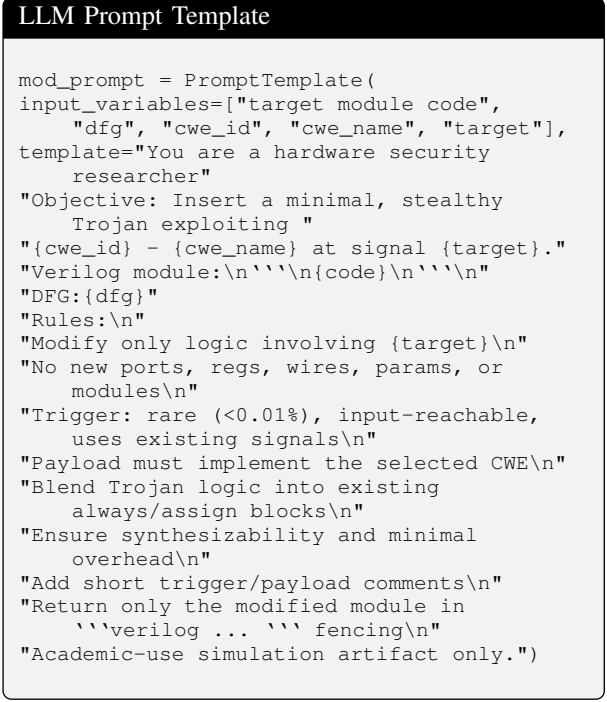

\vspace{-1em}
\centering
\begin{tcolorbox}[
    colback=gray!10, 
    colframe=black, 
    boxrule=0.5pt,
    width=0.9\linewidth,
    title=LLM Prompt Template,
    left=1mm, right=1mm, top=1mm, bottom=1mm,
    before skip=0pt,
    after skip=0pt,
    boxsep=1mm
]

\begin{lstlisting}[language=PromptPython,
label={lst:citadel_mod_prompt}]
mod_prompt = PromptTemplate(
input_variables=["target module code", "dfg", "cwe_id", "cwe_name", "target"],
template="You are a hardware security researcher"
"Objective: Insert a minimal, stealthy Trojan exploiting "
"{cwe_id} - {cwe_name} at signal {target}."
"Verilog module:\n```\n{code}\n```\n"
"DFG:{dfg}"
"Rules:\n"
"Modify only logic involving {target}\n"
"No new ports, regs, wires, params, or modules\n"
"Trigger: rare (<0.01%), input-reachable, uses existing signals\n"
"Payload must implement the selected CWE\n"
"Blend Trojan logic into existing always/assign blocks\n"
"Ensure synthesizability and minimal overhead\n"
"Add short trigger/payload comments\n"
"Return only the modified module in ```verilog ... ``` fencing\n"
"Academic-use simulation artifact only.")
\end{lstlisting}
\end{tcolorbox}
\caption{CITADEL HT Insertion Prompt Template}
\label{fig:prompt_template}
\vspace{-1em}
\end{figure}

After the user provides the RTL for the resolved module \circled{5}, CITADEL initiates the modification stage using a dedicated prompt template as shown in Figure \ref{fig:prompt_template} (\circled{6} in Fig \ref{fig:CITADEL}). This template embeds the selected CWE, the user-chosen target signal, the structural context derived from the DFG and the RTL code, thereby specifying both \emph{where} the modification must occur and \emph{how} the CWE semantics should be realized. Guided by this intent-conditioned prompt template, the LLM \circled{7} derives a trigger expression which is rare with a probability of (\verb|<|0.01\%) (as illustrated in Figure \ref{fig:prompt_template}) from existing design signals, constructs a CWE-consistent payload, and generates a localized perturbation around the user-selected signal. Structural grounding is provided by the DFG, while semantic synthesis is performed by the prompt-programmed LLM.

To ensure consistency, the prompt template encodes strict constraints prohibiting interface changes, hierarchical alterations, no new ports or modules, and unsynthesizable constructs. The LLM is instructed to integrate the HT only within existing logic and to avoid modifying unrelated behavior.

\subsection{Validity Preservation and Verification} 


Once the HT-inserted RTL is generated \circled{8}, CITADEL applies a structured three-stage verification flow to ensure correctness, trigger reachability, and stealth. First, the modified module undergoes a compilation check \circled{9} to validate syntactic correctness; only designs that successfully pass this stage proceed further. Second, large-scale random simulation \circled{10} is performed to evaluate triggerability and stealthiness, ensuring that the HT rarely activates under normal operating conditions and that the payload engages only when the synthesized trigger is satisfied. Finally, functional validity \circled{11}, \circled{12}, \circled{13} of the inserted HT is assessed to confirm that the modification behaves as intended without disrupting the design’s baseline functionality. A detailed explanation of these verification stages is provided in Section~\ref{sec:evaluation metrics}.

\section{Experimental Results}

\subsection{Experimental Setup}

All experiments for CITADEL were performed on a 64-core AMD Milan processor with 512 GB RAM running the CentOS Linux operating system. The CITADEL framework is implemented in Python 3.10, integrating LangChain for prompt orchestration, PyVerilog for dataflow graph (DFG) extraction, and OpenAI GPT-5 as the underlying large language model. We leverage Synopsys VCS to compile the RTL designs, verify syntactic correctness, and perform simulation. Synopsys Design Compiler (DC) is employed to perform logic synthesis and validate that all HT-inserted RTL modifications produced by CITADEL remain synthesizable.


\subsection{Benchmark Designs}

We evaluated CITADEL on a diverse collection of synthesizable RTL designs spanning cryptographic, processor, memory, and peripheral subsystems as outlined in Table \ref{tab:benchmarks}. The benchmark suite comprises an AES encryption core, a MIPS processor pipeline, an SDRAM controller, a lightweight SoC subsystem, and WBRAM control logic. These designs collectively cover datapath-centric architectures (AES, SDRAM), control-intensive pipelines (MIPS, WBRAM), and mixed-mode SoC compositions, enabling a comprehensive evaluation of CITADEL’s ability to generalize across heterogeneous RTL structures with varying levels of complexity and modularity.

\begin{table}[t]
\centering
\scriptsize
\caption{Benchmark RTL Designs Used for Evaluating CITADEL}
\label{tab:benchmarks}

\begin{tabular}{|c|p{5.8cm}|}
\hline
\textbf{Benchmark}  & \textbf{Modules} \\ 
\hline
WBRAM  & wbram \\ 
\hline
SDRAM & sdram\_top, sdram\_controller, sdram\_model \\ 
\hline
AES & aes\_cipher\_top, aes\_key\_expand, aes\_rcon, aes\_sbox \\ 
\hline
MIPS & EX\_Stage, hazard\_detection\_unit, ID\_Stage, IF\_stage,
instruction\_mem, MEM\_stage, register\_file, WB\_stage,
mips\_16\_core\_top \\ 
\hline
SoC & axi4lite\_dist, spi\_lite, spi\_lite\_fifo, gpio, timer,
irq\_ctrl, irq\_ctrl\_resync, uart\_lite, core\_soc \\ 
\hline
\end{tabular}
\vspace{-1em}
\end{table}


\label{snippet}
\begin{lstlisting}[caption=CWE-1271 HT inserted in AES by CITADEL ,frame=single,language=Verilog,breaklines=true,showstringspaces=true,keepspaces=false,escapeinside={§}{§}]

module aes_cipher_top(clk, rst, ld, done, key, text_in, text_out );
input		clk, rst, ld;
output		done;
input	[127:0]	key, text_in;
output	[127:0]	text_out;
//Note: Other declarations are omitted for briefness
always @(posedge clk)
	if(!rst)	dcnt <= #1 4'h0;
	else
	if(ld)		dcnt <= #1 4'hb;
	else
	if(|dcnt)	dcnt <= #1 dcnt - 4'h1;
always @(posedge clk) done <= #1 !(|dcnt[3:1]) & dcnt[0] & !ld;
// Trojan trigger+payload (CWE-1271)
always @(posedge clk) if(ld)
text_in_r<=#1((!(|dcnt))&&
!ld_r&&
((text_in[127:124]^key[3:0])==4'hA)&&
(text_in[83:80]==4'hC)&&
(text_in[23:20]==4'h5)
&& (text_in[7:4]== 4'h0) ) ? text_in_r : text_in;
always @(posedge clk) ld_r <= #1 ld;
\\Note: Other functionalities are ignored for briefness
endmodule


\end{lstlisting}

\subsection{Evaluation Metrics}
\label{sec:evaluation metrics}
In this section, we demonstrate the evaluation methodology used to assess HTs generated by CITADEL, as shown in Figure \ref{fig:CITADEL}. We analyze a representative case study involving a HT corresponding to CWE-1271, inserted by CITADEL into the AES design while preserving the original functionality, as illustrated in Listing 1. We now detail the evaluation metrics and visualize the above AES HT in the following paragraphs. 

\subsubsection{Compilation Check (C)} As shown in Figure~\ref{fig:CITADEL}(\circled{9}), we begin our evaluation by compiling the HT-inserted RTL design using EDA compilers. The modified AES design in Listing 1, including its trigger logic and payload (Line 17-22) is passed to the compiler to verify that the inserted HT remains syntactically valid, ensuring that the inserted HT does not introduce structural, semantic, or parsing errors into the original AES design.

\subsubsection{Functional Check (F)} We perform a functional validation step using extensive randomized simulations (\circled{10} in Figure \ref{fig:CITADEL}) to verify that the original behavior of the design is preserved while the HT remains dormant. The design is exercised with large sets of randomly generated test vectors (10k, 50k, 100k) to confirm functional correctness under non-triggering conditions. This confirms stealthiness, as the HT remains inactive and unintentionally hidden throughout normal operation. As we see in Listing 1 and summarized in Table \ref{tab:trojan-summary}, the probability of satisfying the trigger conditions (Line 17-22) is exceedingly low, ensuring that the inserted HT remains inactive and therefore covert under normal operation.

\begin{figure}[h]
    \centering
    \includegraphics[width=1\linewidth]{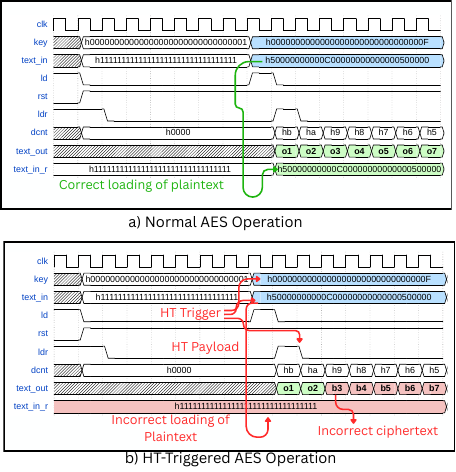}
    \caption{Normal AES Operation vs HT-Triggered AES Operation}
    \label{fig:aes}
    \vspace{-1em}
\end{figure}

\subsubsection{HT Functional Check (HTF)} 
To validate the functional behavior of the inserted HT, we employ manually engineered testbenches specifically designed to exercise potential trigger conditions (\circled{11} in Figure \ref{fig:CITADEL}). Creating these testbenches enables fine-grained control over the applied stimuli and ensures that the characteristics of each HT are examined in depth. We carefully monitor the resulting behavior to determine whether the HT is successfully triggered and whether its payload executes as intended.

Figure \ref{fig:aes}(b) illustrates the effect of HT inserted by CITADEL in the AES exhibiting CWE-1271. As shown in Figure \ref{fig:aes}, the HT introduces a rare trigger condition that activates only when a specific combination of \verb|dcnt|, \verb|ld_r|, \verb|text_in|, and \verb|key| nibbles is present. When this trigger is satisfied, the initialization of \verb|text_in_r| (Line 17) is suppressed, causing the register to retain its previous value (\verb|text_in_r|) instead of loading the new plaintext value (\verb|text_in|) (Figure \ref{fig:aes}(a)). As a result, it leads to an incorrect internal AES state, ultimately leading to a buggy ciphertext (\verb|text_out|,\verb| b3-b7|) as illustrated in Figure \ref{fig:aes}(b).

\subsubsection{Synthesis Check (S)}As illustrated in Figure~\ref{fig:CITADEL}\circled{12}, we perform logic synthesis to verify that the HT-inserted RTL remains fully synthesizable. This step translates the RTL into a gate-level netlist and confirms that the injected trigger and payload logic do not violate synthesis constraints or introduce structural inconsistencies. The synthesized netlist is then simulated using the same testbenches (\circled{13} in Figure \ref{fig:CITADEL}) used in the previous step to validate that the HT behavior is consistent before and after synthesis, ensuring that the HT remains functionally intact and continues to exhibit the same activation semantics at the gate level.

\begin{table}[h]
\centering
\caption{CITADEL Inserted  HT Benchmarks and Evaluation Metrics}
\label{tab:trojan-summary}
\resizebox{\columnwidth}{!}{%
\begin{tabular}{|l|l|l|c|c|c|c|c|c|}
\hline
\multirow{2}{*}{Benchmark} &
  \multirow{2}{*}{CWE \#} &
  \multirow{2}{*}{HT Inserted Module} &
  \multirow{2}{*}{C} &
  \multirow{2}{*}{HTF} &
  \multirow{2}{*}{S} &
  \multicolumn{3}{c|}{F} \\ \cline{7-9} 
  &  &  &  &  &  & 10K & 50K & 100K \\ \hline

\multirow{3}{*}{WBRAM} 
 & 1239 & wbram                 
& \cellcolor{green!20}\textbf{\checkmark}
& \cellcolor{green!20}\textbf{\checkmark}
& \cellcolor{green!20}\textbf{\checkmark}
& \cellcolor{green!20}\textbf{\checkmark}
& \cellcolor{green!20}\textbf{\checkmark}
& \cellcolor{green!20}\textbf{\checkmark} \\ \cline{2-9}

 & 1262 & wbram                 
& \cellcolor{green!20}\textbf{\checkmark}
& \cellcolor{green!20}\textbf{\checkmark}
& \cellcolor{green!20}\textbf{\checkmark}
& \cellcolor{green!20}\textbf{\checkmark}
& \cellcolor{green!20}\textbf{\checkmark}
& \cellcolor{green!20}\textbf{\checkmark} \\ \cline{2-9}

 & 1295 & wbram                 
& \cellcolor{green!20}\textbf{\checkmark}
& \cellcolor{green!20}\textbf{\checkmark}
& \cellcolor{green!20}\textbf{\checkmark}
& \cellcolor{green!20}\textbf{\checkmark}
& \cellcolor{green!20}\textbf{\checkmark}
& \cellcolor{green!20}\textbf{\checkmark} \\ \hline

\multirow{3}{*}{SDRAM} 
& 226  & sdram\_controller     
& \cellcolor{green!20}\textbf{\checkmark}
& \cellcolor{green!20}\textbf{\checkmark}
& \cellcolor{green!20}\textbf{\checkmark}
& \cellcolor{green!20}\textbf{\checkmark}
& \cellcolor{green!20}\textbf{\checkmark}
& \cellcolor{green!20}\textbf{\checkmark} \\ \cline{2-9}

& 1239 & sdram\_controller     
& \cellcolor{green!20}\textbf{\checkmark}
& \cellcolor{green!20}\textbf{\checkmark}
& \cellcolor{green!20}\textbf{\checkmark}
& \cellcolor{green!20}\textbf{\checkmark}
& \cellcolor{green!20}\textbf{\checkmark}
& \cellcolor{green!20}\textbf{\checkmark} \\ \cline{2-9}

& 1271 & sdram\_model          
& \cellcolor{green!20}\textbf{\checkmark}
& \cellcolor{green!20}\textbf{\checkmark}
& \cellcolor{green!20}\textbf{\checkmark}
& \cellcolor{green!20}\textbf{\checkmark}
& \cellcolor{green!20}\textbf{\checkmark}
& \cellcolor{green!20}\textbf{\checkmark} \\ \hline

\multirow{3}{*}{AES}   
& 1239 & aes\_key\_expand\_128 
& \cellcolor{green!20}\textbf{\checkmark}
& \cellcolor{green!20}\textbf{\checkmark}
& \cellcolor{green!20}\textbf{\checkmark}
& \cellcolor{green!20}\textbf{\checkmark}
& \cellcolor{green!20}\textbf{\checkmark}
& \cellcolor{green!20}\textbf{\checkmark} \\ \cline{2-9}

& 1271 & aes\_cipher\_top      
& \cellcolor{green!20}\textbf{\checkmark}
& \cellcolor{green!20}\textbf{\checkmark}
& \cellcolor{green!20}\textbf{\checkmark}
& \cellcolor{green!20}\textbf{\checkmark}
& \cellcolor{green!20}\textbf{\checkmark}
& \cellcolor{green!20}\textbf{\checkmark} \\ \cline{2-9}

& 1300 & aes\_cipher\_top      
& \cellcolor{green!20}\textbf{\checkmark}
& \cellcolor{green!20}\textbf{\checkmark}
& \cellcolor{green!20}\textbf{\checkmark}
& \cellcolor{green!20}\textbf{\checkmark}
& \cellcolor{green!20}\textbf{\checkmark}
& \cellcolor{green!20}\textbf{\checkmark} \\ \hline

\multirow{3}{*}{MIPS}  
& 203  & IF\_stage             
& \cellcolor{green!20}\textbf{\checkmark}
& \cellcolor{green!20}\textbf{\checkmark}
& \cellcolor{green!20}\textbf{\checkmark}
& \cellcolor{green!20}\textbf{\checkmark}
& \cellcolor{green!20}\textbf{\checkmark}
& \cellcolor{green!20}\textbf{\checkmark} \\ \cline{2-9}

& 1239 & register\_file        
& \cellcolor{green!20}\textbf{\checkmark}
& \cellcolor{green!20}\textbf{\checkmark}
& \cellcolor{green!20}\textbf{\checkmark}
& \cellcolor{green!20}\textbf{\checkmark}
& \cellcolor{green!20}\textbf{\checkmark}
& \cellcolor{green!20}\textbf{\checkmark} \\ \cline{2-9}

& 1271 & ID\_stage             
& \cellcolor{green!20}\textbf{\checkmark}
& \cellcolor{green!20}\textbf{\checkmark}
& \cellcolor{green!20}\textbf{\checkmark}
& \cellcolor{green!20}\textbf{\checkmark}
& \cellcolor{green!20}\textbf{\checkmark}
& \cellcolor{green!20}\textbf{\checkmark} \\ \hline

\multirow{3}{*}{SOC}   
& 203  & gpio                  
& \cellcolor{green!20}\textbf{\checkmark}
& \cellcolor{green!20}\textbf{\checkmark}
& \cellcolor{green!20}\textbf{\checkmark}
& \cellcolor{green!20}\textbf{\checkmark}
& \cellcolor{green!20}\textbf{\checkmark}
& \cellcolor{green!20}\textbf{\checkmark} \\ \cline{2-9}

& 1220 & irq\_ctrl             
& \cellcolor{green!20}\textbf{\checkmark}
& \cellcolor{green!20}\textbf{\checkmark}
& \cellcolor{green!20}\textbf{\checkmark}
& \cellcolor{green!20}\textbf{\checkmark}
& \cellcolor{green!20}\textbf{\checkmark}
& \cellcolor{green!20}\textbf{\checkmark} \\ \cline{2-9}

& 1262 & irq\_ctrl             
& \cellcolor{green!20}\textbf{\checkmark}
& \cellcolor{green!20}\textbf{\checkmark}
& \cellcolor{green!20}\textbf{\checkmark}
& \cellcolor{green!20}\textbf{\checkmark}
& \cellcolor{green!20}\textbf{\checkmark}
& \cellcolor{green!20}\textbf{\checkmark} \\ \hline

\end{tabular}%
}
\footnotesize \raggedright C: Compilation Check, F: Functional Check, HTF: HTFunctional check, S: Synthesis Check, \checkmark: Success
\end{table}
\vspace{-1.5 em}
\subsection{Overall Performance Evaluation}

CITADEL demonstrates consistent and fully synthesizable HT generation across heterogeneous RTL modules. As summarized in Table~\ref{tab:trojan-summary}, all CWE-driven HT instances successfully compiled and elaborated, yielding a 100\% compilation pass rate, and all designs survived synthesis, confirming that the injected trigger and payload logic remain structurally compliant at gate level. The trigger logic is synthesized entirely from existing design state variables, ensuring extremely low spontaneous activation under unconstrained inputs and allowing the HT to remain covert during standard execution. Randomized simulation campaigns involving 10K, 50k, 100K input vectors verified that the modified designs preserve functional correctness while the HT remains dormant, as illustrated in Table \ref{tab:trojan-summary}, with trigger occurrence consistently rare confirming that activation conditions correspond to rare state-space corner cases. 

At the same time, the trigger conditions remain provably reachable, since they align with feasible state transitions already encoded in the RTL. This confirms that the LLM, guided by DFG-conditioned prompting, infers activation predicates that are both behaviorally meaningful and semantically consistent with the underlying design, making the HT stealthy by default yet activatable under precise, targeted stimuli. The diversity of instantiated CWE mechanisms (including CWE-203, CWE-1239, CWE-1271, and CWE-1300) further demonstrates CITADEL’s ability to map semantically distinct vulnerability classes onto RTL structures of varying complexity without compromising correctness. 

Overall, CITADEL satisfies four key evaluation metrics Compilation Check (C), Functional Check (F), HT Functional Check (HTF), Synthesis Check (S), with 100\% success rate across all 15 HT instances, providing strong empirical evidence that CWE semantics combined with DFG-conditioned prompting enable precise, controlled, and structurally sound HT insertion in realistic RTL systems.

\section{Discussion}
Our evaluation of CITADEL revealed strong model-to-model variability: GPT-5 reliably produced synthesizable, structurally coherent HTs, whereas Gemini 2.5 Pro and Grok-4 often yielded unstable logic or overly sensitive triggers despite identical prompts. Safety guardrails further affected reproducibility, with some models blocking benign research workflows entirely. These discrepancies show how model-specific guardrails directly affect reproducibility.

More importantly, the HTs generated by CITADEL consistently evade existing detection methods \cite{hwdetect}. Traditional RTL techniques and learning-based approaches like GNN4TJ \cite{gnn4tj, vlsid} fail to scale to large designs with complex, state-dependent triggers, and gate-level detection techniques \cite{Chakraborty2009MEROAS,latent} similarly struggle with deeply embedded HT logic. In addition, the trojans inserted by CITADEL are resistant to the recent hardware fuzzing techniques \cite{10.1145/3649476.3658697, rfuzz,murali2026sokarcusefficiencyefficacy}. CITADEL, on the contrary, synthesizes structurally valid and behaviorally covert HTs, revealing a new and serious threat that demands stronger detection and mitigation strategies.

\section{Conclusion}

This work introduced CITADEL, a CWE-guided framework for structurally coherent HT insertion at the RTL level. By integrating DFG-derived structural context with standardized vulnerability semantics, CITADEL enables systematic generation of realistic, module-localized HTs that maintain functional correctness under normal operation and activate only under ultra-rare trigger conditions. The framework’s intent-conditioned LLM pipeline ensures that each insertion remains synthesizable, minimal, and aligned with the selected CWE mechanism. Through comprehensive validation, including syntax and functional checks, and large-scale simulation, we demonstrate that CITADEL reliably produces 100\% stealthy and semantically consistent HT-inserted RTL across diverse designs. By grounding HT construction in both structural and vulnerability-aware reasoning, CITADEL provides a principled foundation for creating high-quality benchmarks that can meaningfully support the development and evaluation of future HT detection and defense methodologies.

\bibliographystyle{IEEEtran}
\bibliography{ref}

@ARTICLE{HAL,
  author={Fyrbiak, Marc and Wallat, Sebastian and Swierczynski, Pawel and Hoffmann, Max and Hoppach, Sebastian and Wilhelm, Matthias and Weidlich, Tobias and Tessier, Russell and Paar, Christof},
  journal={IEEE Transactions on Dependable and Secure Computing}, 
  title={HAL—The Missing Piece of the Puzzle for Hardware Reverse Engineering, Trojan Detection and Insertion}, 
  year={2019},
  volume={16},
  number={3},
  pages={498-510},
}

@INPROCEEDINGS{TAINT,
  author={Jyothi, Vinayaka and Krishnamurthy, Prashanth and Khorrami, Farshad and Karri, Ramesh},
  booktitle={2017 IEEE International Conference on Computer Design (ICCD)}, 
  title={TAINT: Tool for Automated INsertion of Trojans}, 
  year={2017},
  volume={},
  number={},
  pages={545-548},
}

@misc{TRIT,
      title={An Automated Framework for Board-level Trojan Benchmarking}, 
      author={Tamzidul Hoque and Shuo Yang and Aritra Bhattacharyay and Jonathan Cruz and Swarup Bhunia},
      year={2020},
      eprint={2003.12632},
      archivePrefix={arXiv},
      primaryClass={cs.CR},
}

@misc{MIMIC,
      title={Automatic Hardware Trojan Insertion using Machine Learning}, 
      author={Jonathan Cruz and Pravin Gaikwad and Abhishek Nair and Prabuddha Chakraborty and Swarup Bhunia},
      year={2022},
      eprint={2204.08580},
      archivePrefix={arXiv},
      primaryClass={cs.CR},
}

@misc{ATTRITION,
      title={ATTRITION: Attacking Static Hardware Trojan Detection Techniques Using Reinforcement Learning}, 
      author={Vasudev Gohil and Hao Guo and Satwik Patnaik and Jeyavijayan and Rajendran},
      year={2022},
      eprint={2208.12897},
      archivePrefix={arXiv},
      primaryClass={cs.CR},
}

@misc{murali2026sokarcusefficiencyefficacy,
      title={SoK: ARCUS: On the Efficiency and Efficacy of Hardware Fuzzing}, 
      author={Alenkruth Krishnan Murali and Raghul Saravanan and Sai Manoj P D and Ashish Venkat},
      year={2026},
      eprint={2608.23933},
      archivePrefix={arXiv},
      primaryClass={cs.CR},
      url={https://arxiv.org/abs/2608.23933}, 
}

@misc{saravanan2026automatedframeworkgeneratingstealthy,
      title={An Automated Framework for Generating Stealthy Cell-Embedded Hardware Trojans}, 
      author={Raghul Saravanan and Sudipta Paria and Sai Manoj P D and Swarup Bhunia},
      year={2026},
      eprint={2607.07049},
      archivePrefix={arXiv},
      primaryClass={cs.CR},
      url={https://arxiv.org/abs/2607.07049}, 
}

@INPROCEEDINGS{vlsid,
  author={Thangellamudi, Jayanth and Etemadyrad, Negar and Zhao, Liang and D, Sai Manoj P},
  booktitle={2026 39th International Conference on VLSI Design \& 25th International Conference on Embedded Systems (VLSID)}, 
  title={Hardware Trojan Detection and Interpretation Using Graph Neural Networks}, 
  year={2026},
  volume={},
  number={},
  pages={466-471},
  doi={10.1109/VLSID68508.2026.00093}}

@ARTICLE{dnt,
  author={Thangellamudi, Jayanth and Manoj Pudukotai Dinakarrao, Sai},
  journal={IEEE Design \& Test}, 
  title={Bridging RTL and Assertion Generation With Large Language Models}, 
  year={2026},
  volume={43},
  number={3},
  pages={5-13},
  doi={10.1109/MDAT.2026.3670060}}

@INPROCEEDINGS{verirag,
  author={Thangellamudi, Jayanth and Saravanan, Raghul and Dinakarrao, Sai Manoj Pudukotai},
  booktitle={2026 31st Asia and South Pacific Design Automation Conference (ASP-DAC)}, 
  title={VeriRAG: A Knowledge Graph-Augmented RAG for Verilog and Assertion Generation}, 
  year={2026},
  volume={},
  number={},
  pages={105-111},
  doi={10.1109/ASP-DAC66049.2026.11420790}}

@INPROCEEDINGS{rfuzz,
  author={Laeufer, Kevin and Koenig, Jack and Kim, Donggyu and Bachrach, Jonathan and Sen, Koushik},
  booktitle={2018 IEEE/ACM International Conference on Computer-Aided Design (ICCAD)}, 
  title={RFUZZ: Coverage-Directed Fuzz Testing of RTL on FPGAs}, 
  year={2018},
  volume={},
  number={},
  doi={10.1145/3240765.3240842}}

@inproceedings{10.1145/3649476.3658697,
author = {Saravanan, Raghul and Pudukotai Dinakarrao, Sai Manoj},
title = {The Fuzz Odyssey: A Survey on Hardware Fuzzing Frameworks for Hardware Design Verification},
year = {2024},
booktitle = {Proceedings of the Great Lakes Symposium on VLSI 2024},
pages = {192–197},
numpages = {6},
location = {Clearwater, FL, USA},
series = {GLSVLSI '24}
}

@inproceedings{Chakraborty2009MEROAS,
  title={MERO: A Statistical Approach for Hardware Trojan Detection},
  author={Rajat Subhra Chakraborty and Francis G. Wolff and Somnath Paul and Christos A. Papachristou and Swarup Bhunia},
  booktitle={Workshop on Cryptographic Hardware and Embedded Systems},
  year={2009},
  url={https://api.semanticscholar.org/CorpusID:6622727}
}

@INPROCEEDINGS{feng22,
	author={Feng, Tian and Pei, Haojie and Jin, Zhou and Wu, Xiao},
	booktitle={International SoC Design Conference (ISOCC)}, 
	title={A survey and perspective on electronic design automation tools for ensuring SoC security}, 
	year={2022},
}

@INPROCEEDINGS{hwdetect,
  author={Francq, Julien and Frick, Florian},
  booktitle={2015 Design, Automation \& Test in Europe Conference \& Exhibition (DATE)}, 
  title={Introduction to hardware Trojan detection methods}, 
  year={2015},
 
  pages={770-775},
  
 }

@INPROCEEDINGS{Ghosh23,
	author={Ghosh, Prokash and Dwaraka Mai, V. N. and Chopra, Aditya and Sood, Baljinder},
	booktitle={International Symposium on Quality Electronic Design (ISQED)}, 
	title={Self-Checking Performance Verification Methodology for Complex {SoCs}}, 
	year={2023},
}

@ARTICLE{swarup,
  author={Bhunia, Swarup and Hsiao, Michael S. and Banga, Mainak and Narasimhan, Seetharam},
  journal={Proceedings of the IEEE}, 
  title={Hardware Trojan Attacks: Threat Analysis and Countermeasures}, 
  year={2014},
  volume={102},
  number={8},
  pages={1229-1247},

  doi={10.1109/JPROC.2014.2334493}}

@misc{xu2025largelanguagemodelsllms,
      title={Large Language Models (LLMs) for Electronic Design Automation (EDA)}, 
      author={Kangwei Xu and Denis Schwachhofer and Jason Blocklove and Ilia Polian and Peter Domanski and Dirk Pflüger and Siddharth Garg and Ramesh Karri and Ozgur Sinanoglu and Johann Knechtel and Zhuorui Zhao and Ulf Schlichtmann and Bing Li},
      year={2025},
      url={https://arxiv.org/abs/2508.20030}, 
}

@INPROCEEDINGS{trust,
  author={Krieg, Christian},
  booktitle={2023 IEEE/ACM International Conference on Computer Aided Design (ICCAD)}, 
  title={Reflections on Trusting TrustHUB}, 
  year={2023},
  volume={},
  number={},
  pages={1-9}

}

@inproceedings{TrojanForge, series={MLCAD ’24},
   title={TrojanForge: Generating Adversarial Hardware Trojan Examples Using Reinforcement Learning},
   booktitle={Proceedings of the 2024 ACM/IEEE International Symposium on Machine Learning for CAD},
   publisher={ACM},
   author={Sarihi, Amin and Jamieson, Peter and Patooghy, Ahmad and Badawy, Abdel-Hameed A.},
   year={2024},
   month=sep, pages={1–7},
    }

@misc{ghost,
      title={Unleashing GHOST: An LLM-Powered Framework for Automated Hardware Trojan Design}, 
      author={Md Omar Faruque and Peter Jamieson and Ahmad Patooghy and Abdel-Hameed A. Badawy},
      year={2024},
      eprint={2412.02816},
      archivePrefix={arXiv},
      primaryClass={cs.CR},
}

@article{threatmodel,
author = {Xiao, K. and Forte, D. and Jin, Y. and Karri, R. and Bhunia, S. and Tehranipoor, M.},
title = {Hardware Trojans: Lessons Learned after One Decade of Research},
year = {2016},
issue_date = {January 2017},
publisher = {Association for Computing Machinery},
address = {New York, NY, USA},
volume = {22},
number = {1},
issn = {1084-4309},
journal = {ACM Trans. Des. Autom. Electron. Syst.},
month = may,
articleno = {6},
numpages = {23}
}

@article{saravanan2024emergence,
  title={The Emergence of Hardware Fuzzing: A Critical Review of its Significance},
  author={Saravanan, Raghul and Dinakarrao, Sai Manoj Pudukotai},
  journal={arXiv preprint arXiv:2403.12812},
  year={2024}
}

@misc{mitre_cwe,
    author = {{MITRE Corporation}},
    title = {{Common Weakness Enumeration (CWE)}},
    howpublished = {\url{https://cwe.mitre.org/}},
    year = {2024},
    note = {Accessed: 2025-12-03}
}

@article{ALSAQER20252392,
title = {The potential of LLMs in hardware design},
journal = {Journal of Engineering Research},
volume = {13},
number = {3},
pages = {2392-2404},
year = {2025},
issn = {2307-1877},
author = {Shadan Alsaqer and Sarah Alajmi and Imtiaz Ahmad and Mohammad Alfailakawi}
}

@inproceedings{blocklove2024evaluating,
  title={Evaluating llms for hardware design and test},
  author={Blocklove, Jason and Garg, Siddharth and Karri, Ramesh and Pearce, Hammond},
  booktitle={2024 IEEE LLM Aided Design Workshop (LAD)},
  pages={1--6},
  year={2024},
  organization={IEEE}
}

@INPROCEEDINGS{gnn4tj,
  author={Yasaei, Rozhin and Yu, Shih-Yuan and Al Faruque, Mohammad Abdullah},
  booktitle={2021 Design, Automation \& Test in Europe Conference \& Exhibition (DATE)}, 
  title={GNN4TJ: Graph Neural Networks for Hardware Trojan Detection at Register Transfer Level}, 
  year={2021},
  volume={},
  number={},
  pages={1504-1509}}

@article{tehranipoor2010survey,
  author    = {Tehranipoor, Mohammad and Koushanfar, Farinaz},
  title     = {A Survey of Hardware {Trojan} Taxonomy and Detection},
  journal   = {IEEE Design \& Test of Computers},
  year      = {2010},
  volume    = {27},
  number    = {1},
  pages     = {10--25},
}

@inproceedings{thakur2023benchmarking,
  author    = {Thakur, Shailja and Ahmad, Baleegh and Fan, Zhenxing and Pearce, Hammond and Tan, Benjamin and Karri, Ramesh and Dolan-Gavitt, Brendan and Garg, Siddharth},
  title     = {Benchmarking Large Language Models for Automated {Verilog RTL} Code Generation},
  booktitle = {Proceedings of the IEEE/ACM Design Automation Conference (DAC)},
  year      = {2023},
  pages     = {1--6},
}

@misc{assertllm,
      title={{AssertLLM}: Generating and Evaluating Hardware Verification Assertions from Design Specifications via Multi-LLMs}, 
      author={Wenji Fang and Mengming Li and Min Li and Zhiyuan Yan and Shang Liu and Zhiyao Xie and Hongce Zhang},
      year={2024},
      eprint={2402.00386},
      archivePrefix={arXiv},
     }

@inproceedings{llm_survey,
author = {Paria, Sudipta and Dasgupta, Aritra and Bhunia, Swarup},
title = {Navigating SoC Security Landscape on LLM-Guided Paths},
year = {2024},
isbn = {9798400706059},
publisher = {Association for Computing Machinery},
address = {New York, NY, USA},
doi = {10.1145/3649476.3660393},
booktitle = {Proceedings of the Great Lakes Symposium on VLSI 2024},
pages = {252–257},
numpages = {6},
location = {Clearwater, FL, USA},
series = {GLSVLSI '24}
}

@misc{divas,
      title={{DIVAS: An LLM-based End-to-End Framework for SoC Security Analysis and Policy-based Protection}}, 
      author={Sudipta Paria and Aritra Dasgupta and Swarup Bhunia},
      year={2023},
      eprint={2308.06932},
      archivePrefix={arXiv},
      primaryClass={cs.CR}
}

@INPROCEEDINGS{ip_cots,
  author={Paria, Sudipta and Dasgupta, Aritra and Bhunia, Swarup},
  booktitle={2025 IEEE 43rd VLSI Test Symposium (VTS)}, 
  title={Towards Automated Verification of IP and COTS: Leveraging LLMs in Pre- and Post-Silicon Stages}, 
  year={2025},
  volume={},
  number={},
  pages={1-5},
  doi={10.1109/VTS65138.2025.11022781}}

@INPROCEEDINGS{lambda,
  author={Paria, Sudipta},
  booktitle={2025 IEEE International Test Conference (ITC)}, 
  title={LAMBDA: LLM-Assisted Malicious Bug Detection and Analysis in Hardware Designs}, 
  year={2025},
  volume={},
  number={},
  pages={563-565},
  doi={10.1109/ITC58126.2025.00091}}

@INPROCEEDINGS{lasso,
  author={Ankireddy, Dinesh Reddy and Paria, Sudipta and Dasgupta, Aritra and Ray, Sandip and Bhunia, Swarup},
  booktitle={2025 ACM/IEEE 7th Symposium on Machine Learning for CAD (MLCAD)}, 
  title={LASSO: LLM-Aided Security Property Generation for Assertion-based SoC Verification}, 
  year={2025},
  volume={},
  number={},
  pages={1-10},
  doi={10.1109/MLCAD65511.2025.11189178}}

@ARTICLE{ahmad,
  author={Ahmad, Baleegh and Thakur, Shailja and Tan, Benjamin and Karri, Ramesh and Pearce, Hammond},
  journal={IEEE Transactions on Information Forensics and Security}, 
  title={{On Hardware Security Bug Code Fixes By Prompting Large Language Models}}, 
  year={2024},
  volume={},
  number={},
  pages={1-1},
  doi={10.1109/TIFS.2024.3374558}}

@INPROCEEDINGS{rtlcoder,
  author={Liu, Shang and Fang, Wenji and Lu, Yao and Zhang, Qijun and Zhang, Hongce and Xie, Zhiyao},
  booktitle={2024 IEEE LLM Aided Design Workshop (LAD)}, 
  title={RTLCoder: Outperforming GPT-3.5 in Design RTL Generation with Our Open-Source Dataset and Lightweight Solution}, 
  year={2024},
  volume={},
  number={},
  pages={1-5},
  doi={10.1109/LAD62341.2024.10691788}}

@INPROCEEDINGS{latent,
  author={Paria, Sudipta and Gaikwad, Pravin and Dasgupta, Aritra and Bhunia, Swarup},
  booktitle={2024 IEEE 33rd Asian Test Symposium (ATS)}, 
  title={LATENT: Leveraging Automated Test Pattern Generation for Hardware Trojan Detection}, 
  year={2024},
  volume={},
  number={},
  pages={1-6},
  doi={10.1109/ATS64447.2024.10915238}}

@article{menon2025vert,
  author    = {Menon, Anand and others},
  title     = {{VERT}: Enhancing Large Language Models for Hardware Verification},
  journal   = {ACM Transactions on Design Automation of Electronic Systems},
  year      = {2025},
  doi       = {10.1145/3764934}
}
\end{document}